\documentclass{iaus}
\usepackage{graphicx}

\title[Turbulence in the Magellanic Stream] 
{ Turbulence: A Probe of the Dynamics and Physics of the Magellanic Stream   \\  }

\author[Itzhak Goldman]   
{Itzhak Goldman  
  \thanks{Visiting Researcher, Department of Astronomy and Astrophysics, Tel Aviv University, Tel Aviv, Israel.} 
}
\affiliation{ Afeka - Tel Aviv Academic College of Engineering, Bnei Efraim 218, Tel Aviv 69107, Israel 
 \\ email: {\tt goldman@afeka.ac.il} \\[\affilskip] }

\pubyear{2008}
\volume{256}  
\pagerange{1 --3}
\jname{The Magellanic System: Stars, Gas, and Galaxies}
 \editors{ Jacco van Loon \& Joana Oliveira, eds.}
\begin{document}

\maketitle

\begin{abstract}
  A  recent paper by   Stanimirovi{\'c} \etal \ (2008) presents quit interesting results  from H$_{\rm I}$ 
observations of the Magellanic Stream (MS) tip. The high spatial resolution of the  data    reveals   rich and complex  morphological and   kinematic
structures; notably four coherent $H_I$ substreams   extending over angular size of about $20^o$ were found.

We suggest to use the data to search for the existence  
 of an underlying turbulence in the residuals of velocity fields. If existent, a turbulence  would provide a {\it dynamical} evidence ,that the sub streams are cohorent structures. The characteristics of the turbulence
 could yield information about the energy source, as well as about the physical parameters of the gas in these streams.
  
   We use the position-velocity images of  Stanimirovi{\'c} \etal \ (2008) to derive   spatial power spectra for the velocity residuals. These, indicate 
the presence of a large scale turbulence with  size comparable   to  that of the streams themselves. The turbulent velocity on the largest scale is estimated to be about  $15 \ {\rm km/s}$.
Adopting, a distance of $120 \  {\rm kpc}$, implies a turbulent largest scale of $\sim 40 \  {\rm kpc}$ and timescale for decay of about  $3 {\ \rm Gyr}$. 

For a turbulence with scale that large, the natural  energy source   is the tidal interaction between the Magellanic Clouds, and between them and the Milky Way galaxy. The  estimated turbulent timescale for decay is consistent with this mechanism. Such a mechanism has been suggested for the turbulence in the ISM of the SMC by Goldman (2000, 2007). In effect, the turbulence is a fossil from the era of the streams formation.

The shape of derived turbulence spectrum is used here  to  obtain constraints on the inclination of the streams and on the density of the emitting neutral hydrogen.

\end{abstract}

 \section{Introduction}
In a recent paper    Stanimirovi{\'c} \etal \ (2008) present  quit interesting   results  from H$_{\rm I}$ observations of the tip of the Magellanic Stream (MS) . The high spatial resolution of the data obtained by   Stanimirovi{\'c} \etal \ (2008), reveals   rich and complex  morphological and kinematic structures. The authors find four coherent $H_I$ substreams in the tip of the MS extending over projected angular size of about $20^o$. Three of the 
these streams (S2, S3, S4) originate from about the same location and are clumpy. 
The remaining, S1 stream, seems more diffuse and doesn't share the common location of the
 former streams. In all streams, the kinematic data show  large scale velocity gradients   of $\sim  (5-10){\rm km\ s^{-1}\ deg^{-1}}$.

By comparing the observations to the simulations of Connors \etal \ (2006),  Stanimirovi{\'c} \etal \ (2008), interpret the three former streams to be the result of the tidal splitting
 of the main MS  by   tidal interaction of the LMC and the MS about  1.05 Gyr  and  0.55 Gyr ago. In this picture, the MS stream itself was formed  about 1.5  Gyr ago by close tidal encounter of the SMC, LMC and the Milky Way (MW).
  
The   S1 sub-stream, is interpreted to have formed much more recently, about  0.2 Gyr ago, and consists of gas drawn from the Magellanic Bridge. Contrary to the former three streams, it had not enough  time to cool and fragment. 

\section{Present Work}

The tidal interactions assumed to create the streams, generate large scale shear flows, as indeed is evident in the data of  Stanimirovi{\'c} \etal \ (2008). These, in turn, are bound to create turbulence in the ISM by   several instabilities. Turbulence can be created also by shocks via the Richtmeir-Meshkov  instability. The ultimate energy source for all these
instabilities are  the tidal interactions. The result is a large scale turbulence of size
comparable to the entire size of the system. 
Furtheremore, if the decay time of the turbulence turns out to exceed   the
age of the system, the turbulence can serve as a "fossil evidence" that can supply valuable information.  For more details on these issues see Goldman (2000, 2007). 

In the present work we analyze the position-velocity images along the streams, derived by  
 Stanimirovi{\'c} \etal \ (2008). For  each stream, we fitted a  mean velocity field consisting of a constant plus a gradient. This mean velocity field was   substracted from the observed velocity, yielding the residual velocity field, as function of projected angle.  

  We derived the spatial power spectrum for the residual velocity field of each stream. 
The results are shown in Figure 1. The power spectrum for each stream is {\it  a power law} that indicates the presence of an underlying statistical order that reflects   correlations between the fluctuations on different scales. Such a power law is a signature
of the {\it inertial  range} of wavenumbers of a turbulent velocity field.

The turbulence for all  streams encompasses the total extent of the the stream. Such a large scale turbulence can not be generated by a localized source such as supernovae winds. A large scale source with scale at least at that of the stream is required. Tidal interactions is the natural candidate.

The residual velocity fields originate from integration along the line of sight. For scales in the plane of the sky, that are {\it large} compared to the depth along the line of sight, the 
index of the power spectrum power law, equals the index of the turbulent energy spectral function. For scales in the plane of the sky, that are {\it small} compared to the depth along the line of sight, the index of the power spectrum power law, equals the index of the turbulent energy spectral function minus 1.
  
For S2 and S4, the power spectrum can be fitted by a single power law with index $\sim -3$.
This corresponds to a turbulence energy spectral function $E(k)\propto k^{-2}$ which is the inertial range for compressible turbulence, when the depth over which the velocity was integrated, is
larger than the scale in the plane of the sky.

For S1 and S3 there is an indication that  the power spectrum changes from a power law 
with index $\sim -2$ to a power law with index $\sim -3$ at a relative wavenumber of  $\sim 4-5$ . This implies that the depth of these streams is about $4^0$. 
The r.m.s turbulent velocity for all the streams is similar: $\sim 15\ {\rm km s^{-1}}$ namely mildly supersonic. This is consistent with the $-2$ index of the turbulence   energy spectral function.

 \newpage
\section{Discussion}

The main results of the present work are:
\renewcommand{\labelenumi}{\arabic{enumi}.}
\begin{enumerate}
\item 
  Each of the streams exhibits a large scale turbulence   comparable to the size of the stream. This provides {\it dynamical} proof that
  the streams are coherent structures, as stated in  Stanimirovi{\'c} \etal \ (2008).
 \item For an assumed distance of 120 kpc to the MS tip, the projected size of the stream and the largest turbulence scale is $\sim 40$ kpc.
 \item The turbulent r.m.s velocity on the largest scale is about $15\ {\rm km s^{-1}}$, namely mildly supersonic. This is in line with the inertial range index of the turbulent spectral energy function  being $-2$ 
 instead of the Kolmogorov value $-5/3$, appropriate for subsonic turbulence. 
 \item The resulting timescale for decay of the turbulence is about 3 Gyr and is longer than the age of the MS, thus providing "fossil evidence". 
 \item The fact that a break in the power spectrum is evident in S1, and even more so in S3
 indicates that the inclination angle of these stream, with respect to the line of sight is not large; they are viewed almost face-on.
 \item The absence of a clear break in the power spectra of S2 and S4, suggest that their inclination is larger. Taking the absence of the break to imply that the  projected depth is larger than about $8^0$ and assuming that the true depth is $4^0$, as deduced for S2 and S4,
 results in an inclination of $\sim 60 ^0$. From Fig. 6 of Connors \etal \ (2006) one can deduce a similar inclination at the tip of the MS. 
  \item A depth of  $4^0$ at a distance of 120 kpc corresponds to a physical depth of a
8 kpc. For   column densities in the range  $(1-10) \times 10^{19} {\rm cm^{-2}}$ this implies a depth-averaged   density of the warm neutral medium: $n_{WNM} = (0.3 -3) \times 10^{-3} {\rm cm^{-3}}$.
\end{enumerate} 

 \begin{figure}[h]
 \centerline{
\includegraphics[scale=0.75]{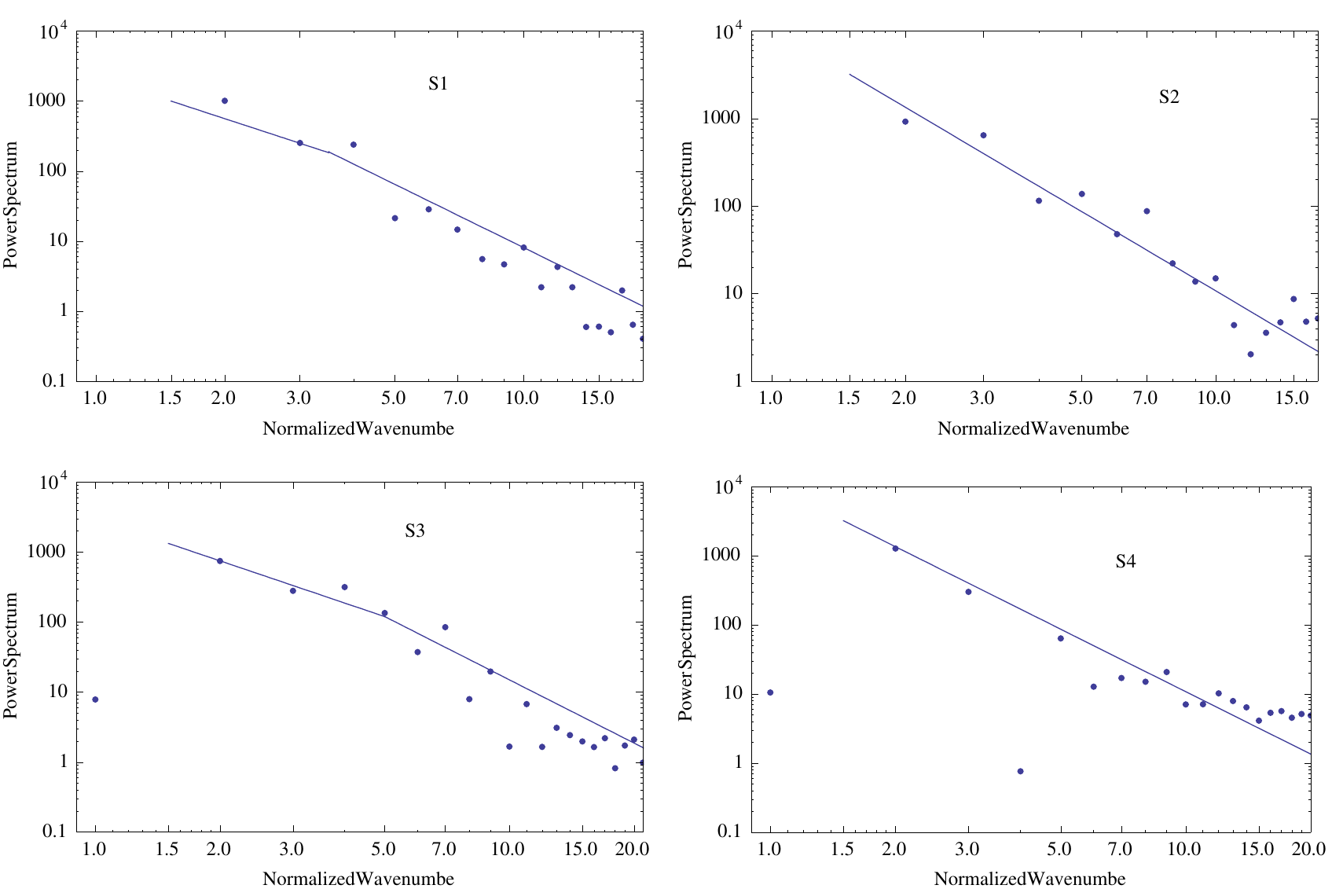}}

\caption{ {\it Dots}: The power spectrum of the  velocity residuals in arbitrary units as function of the normalized wavenumber. Wavenumber 1 corresponds to a largest angular scale in each stream--  S1: $15^0$, S2: $11.9^0$, S3: $17^0$, S4: $15.9^0$.  {\it Lines}:  $k^{-2}, k^{-3}$ for S1 and S3; $k^{-3}$ for S2 and S4.}
 \end{figure}
  
\acknowledgements 

  I would like to thank the conference organizers and the IAU for an IAU award.

\end{document}